\documentclass[lettersize,journal]{IEEEtran}
        \pdfoutput=1
\usepackage{amsfonts}
\usepackage{algorithmic}
\usepackage{algorithm}
\usepackage[caption=false,font=normalsize,labelfont=sf,textfont=sf]{subfig}
\usepackage{textcomp}
\usepackage{stfloats}
\usepackage{url}
\usepackage{verbatim}
\usepackage{cite}

\usepackage{graphicx,wrapfig,amssymb,amsfonts,amsmath,epstopdf,array}
\usepackage[dvipsnames]{xcolor}
\usepackage{tabularx}
\usepackage{array}
\usepackage{xcolor}
\usepackage{tabularray}
\usepackage{multirow}
\usepackage{ragged2e}
\usepackage{vcell}
\usepackage{soul}

\begin{document}

\title{Data-driven Integrated Sensing and Communication: Recent Advances, Challenges, and Future Prospects}

\author{Hammam~Salem,
        MD Muzakkir~Quamar,
        Adeb~Mansoor, Mohammed~Elrashidy, Nasir~Saeed~\IEEEmembership{Senior~Member,~IEEE,} Mudassir Masood~\IEEEmembership{Member,~IEEE,}
}



\maketitle

\begin{abstract}
Integrated Sensing and Communication (ISAC), combined with data-driven approaches, has emerged as a highly significant field, garnering considerable attention from academia and industry. Its potential to enable wide-scale applications in the future sixth-generation (6G) networks has led to extensive recent research efforts. Machine learning (ML) techniques, including $K$-nearest neighbors (KNN), support vector machines (SVM), deep learning (DL) architectures, and reinforcement learning (RL) algorithms, have been deployed to address various design aspects of ISAC and its diverse applications. Therefore, this paper aims to explore integrating various ML techniques into ISAC systems, covering various applications. These applications span intelligent vehicular networks, encompassing unmanned aerial vehicles (UAVs) and autonomous cars, as well as radar applications, localization and tracking, millimeter wave (mmWave) and Terahertz (THz) communication, and beamforming. 
The contributions of this paper lie in its comprehensive survey of ML-based works in the ISAC domain and its identification of challenges and future research directions. By synthesizing the existing knowledge and proposing new research avenues, this survey serves as a valuable resource for researchers, practitioners, and stakeholders involved in advancing the capabilities of ISAC systems in the context of 6G networks.
\end{abstract}


\section{Introduction}
\IEEEPARstart{I}{ntegrated} Sensing and Communication (ISAC) systems hold immense promise in revolutionizing the performance of next-generation wireless networks. By enabling the seamless sharing of critical resources like time, frequency, waveform design, and hardware, ISAC empowers a wide range of applications in 6G that merge sensing and communication systems \cite{tan2021integrated}. Its impact on future wireless systems cannot be overstated, as it perfectly caters to the demands of various domains, such as precise localization, advanced tracking, gesture recognition, activity monitoring, and augmented human reality \cite{tan2021integrated}. With its ability to deliver both high data rates for communication and unparalleled accuracy for sensing, ISAC emerges as an indispensable force driving the future of wireless technology \cite{mu2021integrated}.

Moreover, the increasing trend of using machine learning (ML) techniques in next-generation communication systems aim to improve signal processing, spectrum management, predictive maintenance, and security features. In particular, using deep learning (DL) in ISAC systems aims to reduce hardware dependency, minimize complexity, and overcome the difficulties imposed on the system's respective tasks by the dynamic environment. Therefore, data-driven approaches are expected to enhance automation and efficiency in the overall operation of ISAC systems \cite{mateos2022model, demirhan2023integrated}.

Nevertheless, the implementation of ISAC systems presents several challenges. One such challenge is developing novel waveform designs that can offer the right balance between the requirements of communication and sensing functionalities \cite{chu2022ai}. Beamforming also poses a significant challenge for ISAC systems, given the conflicting preferences of communication and sensing systems. While communication systems favor narrowband configurations to optimize transmission between transmitters and receiver terminals, sensing systems thrive on wideband setups to capture extensive environmental information \cite{demirhan2023integrated}.
Fortunately, researchers have proposed various approaches to tackle these challenges. For instance, in the realm of localization enhancement, a distributed inference framework leveraging device-to-device (D2D) communication was put forth \cite{lee20226g}. Another strategy employed full-duplex operation to address the design intricacies of ISAC systems \cite{barneto2021full}. To achieve simultaneous communication and sensing, the authors in \cite{buzzi2019using} utilized massive multi-input multi-output (MIMO) arrays, enabling communication with moving units while conducting environmental sensing. Additionally, orthogonal frequency-division multiplexing (OFDM) emerged as a viable solution for merging communication and sensing systems, as discussed in \cite{wu2022joint}.

Although these classical methods have made notable strides in improving ISAC system performance, they often encounter a common drawback: the inherent complexity arising from reconciling the distinct requirements of communication and sensing tasks. This complexity can hinder the seamless integration and optimization of these two functionalities. However,  the rapid advancement of ML techniques can provide solutions to mitigating such challenges in ISAC systems. These ML-based solutions have the potential to revolutionize the field, rendering futuristic ISAC technologies more secure, reliable, and efficient. By unlocking the capabilities of ML, we can unlock new dimensions of efficiency, reliability, and security in ISAC systems. Therefore, this paper highlights the significant contributions of ML algorithms in tackling challenges within the realm of ISAC. By conducting a comprehensive review of the literature, this study aims to present a state-of-the-art analysis of the diverse applications of ML-ISAC. These applications encompass various domains, such as vehicles, THz (terahertz) systems, radar technology, beamforming, tracking and localization, spectrum sensing, edge computing, communication environment, intelligent reflective surfaces (IRS), and more. Through this exploration, a comprehensive understanding of the utilization of ML algorithms in ISAC can be obtained, shedding light on the advancements and potential future directions in this field.

\subsection{Relevant Surveys} 

Given that ISAC is an emerging field, several surveys in the literature have approached the topic from different perspectives. As an example, Shao et al. \cite{shao2022machine} presented a comprehensive survey on sensing techniques based on channel state information (CSI) in wireless local access network environments. The authors investigated three categories of techniques: model-based, data-based, and hybrid model-data techniques. They delved into data-based techniques in particular, discussing pattern recognition and deep learning algorithms in detail.
Liu \textit{et al.} \cite{liu2022survey} focused on performance metrics related to communication, sensing, and ISAC, utilizing estimation and information theories. They examined these metrics at individual and joint levels, discussing the limitations of device-based and device-free sensing algorithms. Zhang \textit{et al.} \cite{zhang2021overview} surveyed new signal-processing techniques for integrating communication and radar sensing by exploring technologies employed for communication-centric, radar-centric, and joint optimization and design for joint radar and communication (JRC) systems. In \cite{thoma2021joint}, JRC was surveyed from a waveform and network architecture perspective, discussing waveform generation, processing, and network design.

Furthermore, specific applications of ISAC have also garnered attention in the literature. For instance, Zhang \textit{et al.} \cite{zhang2021enabling} provided insights into the methodologies, challenges, and future directions of applying ISAC in mobile network systems, focusing on signal processing aspects. He \textit{et al.} \cite{he2022beyond} surveyed works that employed IRS in joint localization and communication (JLC), a particular case of ISAC. They explored use cases in communication, localization, and radar-like sensing, emphasizing channel estimation (CE) in IRS-assisted JLC networks. The surveyed works fell into three categories: beam alignment, compressive sensing, and data-driven approaches. Despite these valuable surveys, a comprehensive review specifically focused on the existing solutions of ML algorithms used in ISAC literature needs to be improved. We compare our work against other works in terms of many aspects as shown in table \ref{tab.comp}.  

\begin{table}[h!]
\centering
    \caption{ Comparing this work against other works in terms of the following items: 1) DL/AI roles, 2) ISAC, 3) AVs application, 4) THz application 5) Radar application 6) Beamforming application 7) localization application, 8) spectrum sensing, 9)edge computing, 10) IRS application, 11) channel estimation.} 

\begin{tabular}[h]{|m{1.5em}|m{1em}|m{1em}|m{1em}|m{1em}|m{1em}|m{1em}|m{1em}|m{1em}|m{1em}|m{1em}|m{1em}|} 
\hline
  Ref & 1 & 2 & 3 & 4 & 5 & 6 & 7 & 8 & 9 & 10 & 11\\
\hline
\cite{shao2022machine} & \checkmark & \checkmark & - & - & \checkmark & \checkmark & -& - & - &- & - \\
\hline
\cite{liu2022survey}& - & \checkmark & \checkmark & - & \checkmark & \checkmark & \checkmark & \checkmark & - & \checkmark &  \checkmark \\
\hline
\cite{zhang2021overview} & - & \checkmark & \checkmark & - & \checkmark & \checkmark & - & -  & - & - & \checkmark\\
\hline
\cite{thoma2021joint} & - & \checkmark & - &- & \checkmark &  - & \checkmark & - & - & - & - \\
\hline
\cite{zhang2021enabling} & \checkmark & \checkmark & \checkmark & - & \checkmark & \checkmark & \checkmark & - &  \checkmark & - & \checkmark \\
\hline
\cite{he2022beyond} & \checkmark & \checkmark &  & & & \checkmark & \checkmark & & & \checkmark & \\ 
\hline
\cite{dahrouj2021overview} & \checkmark & - &\checkmark & \checkmark & - & \checkmark & - &- & \checkmark & \checkmark &  \checkmark \\
\hline
this work & \checkmark & \checkmark & \checkmark & \checkmark & \checkmark & \checkmark & \checkmark & \checkmark & \checkmark & \checkmark & \checkmark \\
\hline
    \end{tabular}
    \label{tab.comp}
\end{table}

\subsection{Contributions of this Survey}
Despite the above-discussed surveys, a notable gap exists for an in-depth and comprehensive review dedicated explicitly to exploring the existing solutions of ML algorithms utilized in ISAC. This gap represents a critical opportunity to delve into the realm of ML applications within the context of ISAC, uncovering novel methodologies,  advancements, and key insights.
By bridging this gap, this article explores how ML algorithms have been harnessed in the literature to tackle the challenges associated with ISAC. The primary objective of this study is to present a state-of-the-art review that showcases the diverse range of use cases where ML-ISAC has been successfully employed. These use cases encompass various applications, including vehicular networks, THz systems, radar technology, beamforming, tracking and localization, spectrum sensing, edge computing, communication environment, IRS, and many others. By encompassing these various domains, this work offers a comprehensive overview of the extensive utilization and advancements of ML algorithms within the realm of ISAC, shedding light on the transformative potential of this technology.
Such an endeavor will contribute to academic knowledge and serve as a valuable resource for researchers, practitioners, and industry professionals seeking to navigate the complex intersection of ML and ISAC systems.


\subsection{Organization}
\begin{figure*}
\centering
\includegraphics[width=\textwidth,height=6cm]{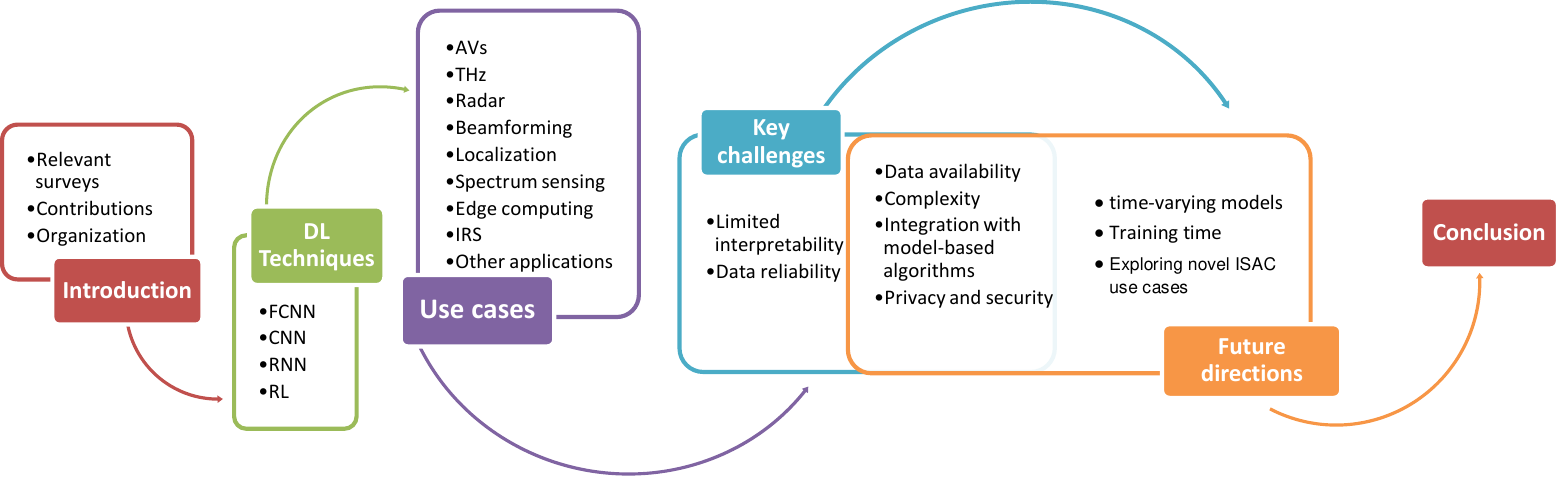}
\caption{Overview diagram of the paper's structure}
\label{Overview}
\end{figure*}

The rest of the paper is organized as follows (c.f., Fig. \ref{Overview}). Section \ref{section:ML} provides a technical background about the mostly used ML algorithms in the literature. Then, use cases for ISAC are discussed in section \ref{section:usecases}. Section \ref{section:challenges} presents the key challenges in ML-ISAC. Finally, future directions and conclusions are presented in sections \ref{section:future} and \ref{section:conc}, respectively. Table \ref{table:abrev} includes the important abbreviations used in this work.



\begin{table}
\label{table:abrev}
\caption{List of key abbreviations.}
\centering
\resizebox{\linewidth}{!}{%
\begin{tabular}{|>{\hspace{0pt}}m{0.104\linewidth}|>{\hspace{0pt}}m{0.327\linewidth}|>{\hspace{0pt}}m{0.104\linewidth}|>{\hspace{0pt}}m{0.404\linewidth}|} 
\hline
\multicolumn{1}{|>{\centering\hspace{0pt}}m{0.104\linewidth}|}{Acronym} & \multicolumn{1}{>{\centering\hspace{0pt}}m{0.327\linewidth}|}{Definition} & \multicolumn{1}{>{\centering\hspace{0pt}}m{0.104\linewidth}|}{Acronym} & \multicolumn{1}{>{\centering\arraybackslash\hspace{0pt}}m{0.404\linewidth}|}{Definition}  \\ 
\hline
6G                                                                      & Sixth generation                                                          & IRS                                                                    & Intelligent reflective surface                                                            \\ 
\hline
AI                                                                      & Artificial intelligence                                                   & ISAC                                                                   & Integrated sensing and communication                                                      \\ 
\hline
APS                                                                     & Azimuth power spectrum                                                    & JRC                                                                    & Joint radar and communication                                                             \\ 
\hline
ASPP                                                                    & Atrous spatial pyramid pooling                                            & KNN                                                                    & K-nearest neighbors                                                                       \\ 
\hline
AV                                                                      & Autonomous vehicles                                                       & LSTM                                                                   & Long short-term memory                                                                    \\ 
\hline
CE                                                                      & Channel estimation                                                        & ML                                                                     & Machine learning                                                                          \\ 
\hline
CNN                                                                     & Convolutional neural network                                              & R2C                                                                    & Radar to communication                                                                    \\ 
\hline
CSI                                                                     & Channel state information                                                 & RL                                                                     & Reinforcement learning                                                                    \\ 
\hline
DDQN                                                                    & Double deep Q-network                                                     & RNN                                                                    & Recurrent neural network                                                                  \\ 
\hline
DL                                                                      & Deep learning                                                             & RSU                                                                    & Roadside unit                                                                             \\ 
\hline
DNN                                                                     & Deep neural network                                                       & SS                                                                     & Spectrum sensing                                                                          \\ 
\hline
DQN                                                                     & Deep Q-network                                                            & SVM                                                                    & Support vector machine                                                                    \\ 
\hline
DRL                                                                     & Deep reinforcement learning                                               & UAV                                                                    & Unmanned aerial vehicle                                                                   \\ 
\hline
EC                                                                      & Edge computing                                                            & UE                                                                     & User equipment                                                                            \\ 
\hline
FCNN                                                                    & Fully connected neural network                                            & V2X                                                                    & Vehicle to everything                                                                     \\ 
\hline
GNN                                                                     & Graph neural network                                                      & V2I                                                                    & Vehicle to infrastructure                                                                 \\
\hline
\end{tabular}
}
\end{table}

\section{Summarizing deep learning techniques}
\label{section:ML}

Deep learning (DL) is a subset of machine learning that uses artificial neural networks with multiple layers. These networks, typically known as deep neural networks (DNNs) \cite{sze2017efficient, miikkulainen2019evolving}, can learn intricate features from data and generate accurate predictions. DL has revolutionized artificial intelligence (AI) and has transformative applications in areas such as computer vision, speech recognition, and autonomous vehicles  \cite{8951131}, communications \cite{9274307}, and many more.

DL techniques have significantly influenced communication systems, leading to breakthroughs in various crucial applications. In the realm of communication, DL has been instrumental in channel coding \cite{CC}, modulation recognition \cite{MR}, beamforming \cite{9845394}, resource allocation \cite{RA}, signal processing \cite{9770266}, and channel estimation \cite{CE}. These applications have leveraged DL's ability to optimize performance, enhance system efficiency, and effectively address the complexities of communication tasks. Notably, DL techniques have gained considerable traction in the context of integrated sensing and communication systems, especially within the framework of 6G networks and terahertz (THz) technologies, where they offer promising avenues for innovation and improvement. In the following, we briefly introduce major DL algorithms used in communication and sensing systems.

\begin{figure}[t]
\centering
\includegraphics[scale=.5]{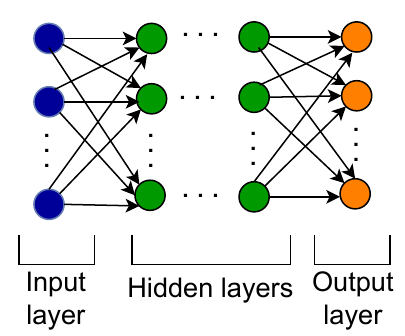}
\caption{ A schematic diagram of FCNN.}
\label{DNN}
\end{figure}

\textbf{Fully Connected Neural Networks (FCNNs):} FCNNs, inspired by the structure and function of the human brain, are neural networks (NNs) that comprise interconnected nodes that represent data features via learnable weights and biases. FCNNs work by processing data through a series of layers, as shown in Fig. \ref{DNN}, where each layer contains a set of nodes. Data is received first by the input layer, and conveyed to the hidden layers for further feature extraction. The inputs at each layer are combined linearly and passed through an activation, non-linear, function. The process is repeated depending on the number of hidden layers. The final predictions are developed at the output layer. FCNNs can be used for many applications, including classification and regression \cite{muller1995neural, abiodun2018state} .

\begin{figure}[t]
\centering
\includegraphics[width=85mm]{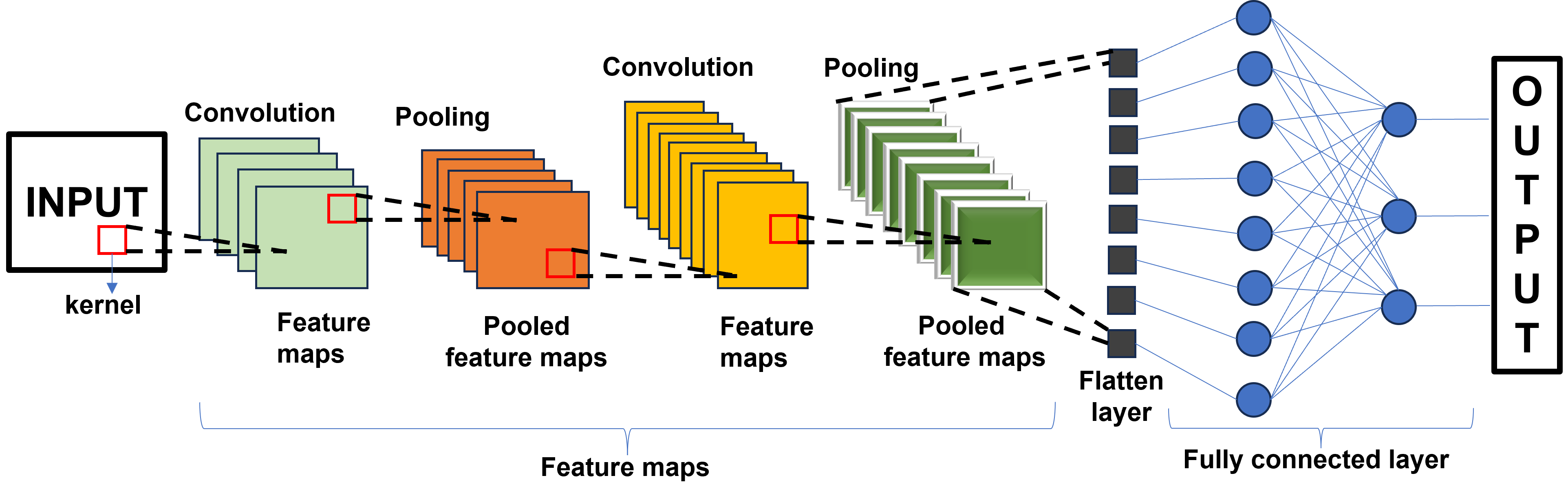}
\caption{A schematic diagram of CNN.}
\label{CNN3}
\end{figure}

\textbf{Convolutional Neural Networks (CNNs):} A CNN is an NN that is used for a variety of tasks, particularly when spatial structure of the data is considered. CNNs consist of layers that perform 2D or 1D convolution, where the convolving masks are learnable parameters. Usually, a convolutional layer is followed by pooling layers, batch normalization layers, and activation layers. The 2D output of the last convolutional block is usually flattened and passed through fully-connected layers to get the final output \cite{li2021survey, o2015introduction}. A example CNN is shown in Fig. \ref{CNN3}.
    
\begin{figure}[t]
\centering
\includegraphics[scale=.3]{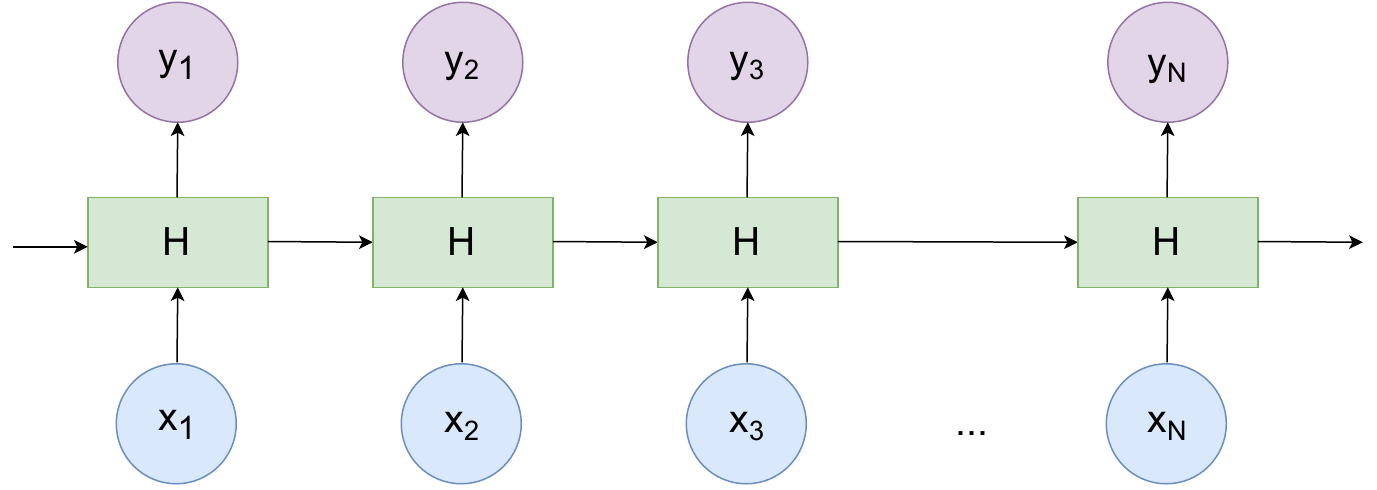}
\caption{ A schematic diagram of RNN.}
\label{RNN}
\end{figure}

 \textbf{Recurrent Neural Networks (RNNs):} RNNs are NNs that consideres the sequential structure of the data (e.g., time sequences). A general scheme for RNNs is shown in Fig. \ref{RNN}. RNNs consist of sequential units, each of which can be represented mathematically by a state. The state depends on the input data and the information received from the previous state. While feedforward networks (e.g., FCNNs and CNNs) assign different parameters for each layer, RNNs share  the same parameters between the states. There are various types of RNN units such as gated recurrent units (GRUs) \cite{dey2017gate} and long short-term memory (LSTM) units \cite{sherstinsky2020fundamentals, yu2019review}. 

\begin{figure}[t]
\centering
\includegraphics[width=90mm]{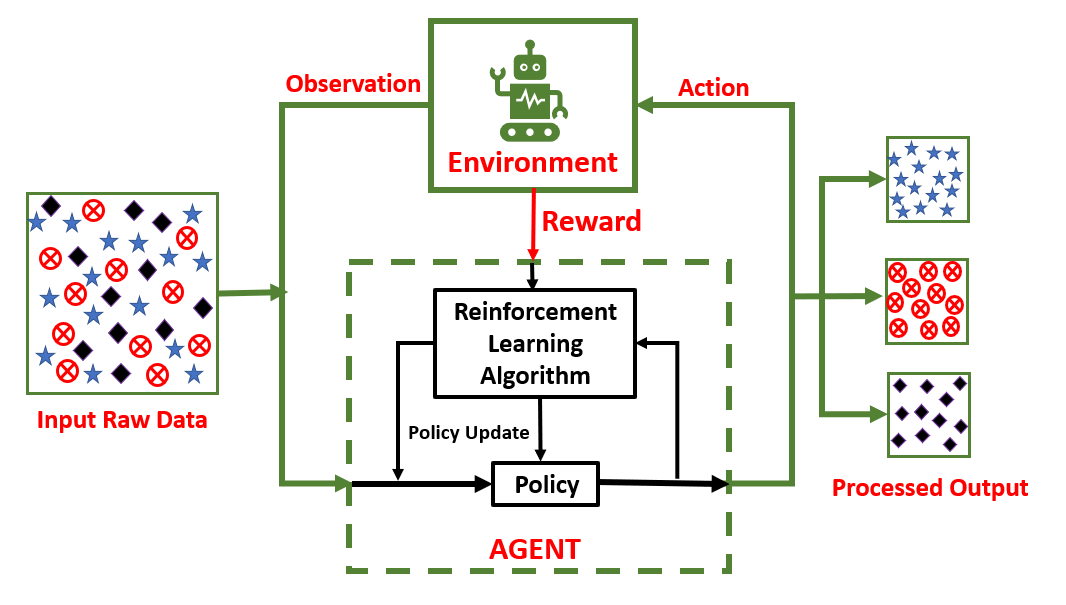}
\caption{ A schematic diagram of RL.}
\label{RL}
\end{figure}

\textbf{Reinforcement Learning (RL):} RL is a sub-field of ML that involves training an agent to make decisions based on observations and feedbacks. The goal is to maximize a reward scalar function, which requires the agent to learn a policy that maps environment states to actions. The RL framework comprises three main components: the agent, the environment, and the reward. The agent observes the environment's state, chooses an action based on its policy, receives a reward, and updates its policy based on the reward and the new state. This cycle continues until the agent achieves the optimal policy or meets the stopping criterion \cite{sutton2018reinforcement, franccois2018introduction}. Fig. \ref{RL} illustrates the basic RL framework.
    
    
The reader is assumed hereafter to understand the main function of each mentioned ML technique. The next section discusses the vital role of the ML algorithms in many use cases of ISAC.

\section{Use cases of Data-driven ISAC systems}
\label{section:usecases}
This section discusses different applications and system settings where ISAC and ML techniques were employed in the literature. It is evident that many works had an overlap of different use cases. Therefore, we emphasize the works done for every use case in their respective sections and refer to the other works that adopted a certain use case as a part of their complete framework for comprehensiveness.

\subsection{Data-driven approached for ISAC-assisted autonomous vehicular networks} 
\begin{figure}[t]
\centering
\includegraphics[width=90mm]{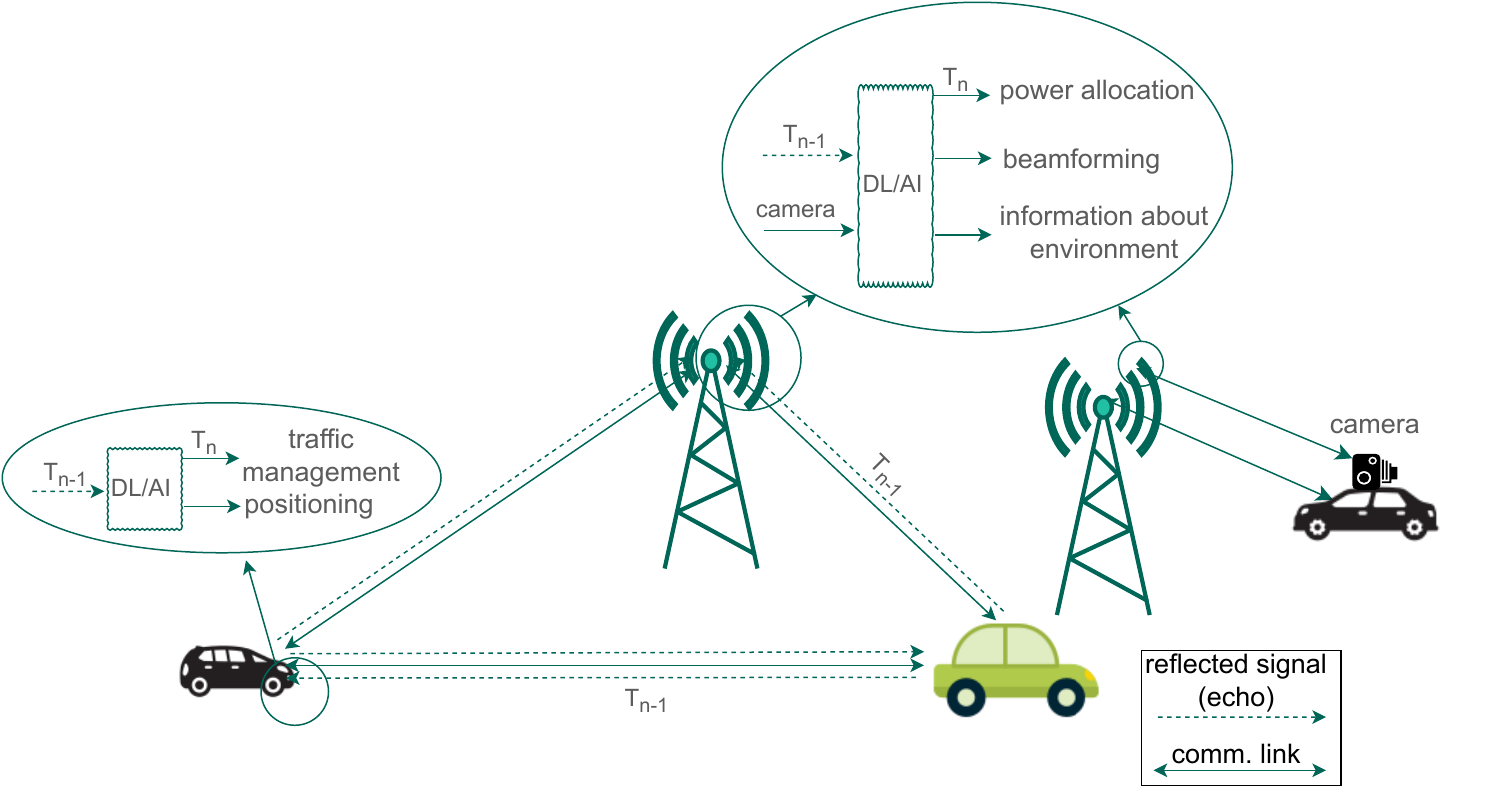}
\caption{An illustration of a vehicular network that utilizes ISAC where: 1) A BS communicates with vehicles and uses the reflected signal for sensing purposes. 2) Two vehicles communicate with each other and utilize the reflected signal for traffic management.}
\label{AVs}
\end{figure}

The robustness of autonomous vehicular networks using ISAC technology will be improved with 6G technology. An example vehicular network is depicted in Fig. \ref{AVs}, where an ML-powered base station and two autonomous vehicles (AVs) extract sensing information from reflected signals. The base station uses ML to extract environment-related data and insights about nearby locations, other AVs, power allocation, and beamforming. The two AVs also communicate using reflected signals, extracting each other's locations and detecting obstacles using ML algorithms.

Waveform optimization is a significant research area in ISAC-assisted AV networks. The preamble data in ISAC packet-based transmission systems can serve as a sensing source. However, a trade-off exists between increasing it to improve sensing accuracy and reducing it to maximize data rate. This trade-off complicates ISAC signal design for AV networks. To tackle this challenge, Chu \textit{et al.} \cite{chu2022ai} proposed an RL-based approach that uses channel quality and maximum packet information as a state input, rather than relying on complete surroundings knowledge. The authors used two algorithms, a Q-table-based method and a deep RL (DRL) approach using an FCNN, to discover optimal signal design in dynamic channel environments. Zhang \textit{et al.} \cite{zhang2023deep1} directed their attention to ISAC vehicular networks where an independent system communicates with multiple users while  detecting multiple vehicular targets simultaneously. The authors proposed ISAC signal schemes based on OFDM and Non-Orthogonal Multiple Access (NOMA). For the communication task, they trained a deep neural network (DNN) for demodulation, and for tracking, they employed the YOLOv5-SORT algorithm.
The studies in \cite{chu2022ai, zhang2023deep1} contribute to advancing ISAC-assisted AV networks by addressing signal design challenges and exploring the potential of RL and DL techniques in optimizing communication and sensing capabilities.

Several other studies have explored DL approaches for predictive beamforming and function selection in ISAC-assisted AVs. For instance, in \cite{mu2021integrated}, Mu \textit{et al.} designed and trained an FCNN to predict the relative angles of the $k$-th vehicle concerning a Road Side Unit (RSU) at a given time $n$. The input to the DL model consisted of received echoes from the target vehicles, and the estimated angular parameters was utilized to design an appropriate beamformer. Then, the authors in \cite{hieu2022transferable} focused on ISAC-assisted AVs and proposed a function selection process to determine whether communicate or sense should take precedence at a specific time slot. Markov Decision Process (MDP) was employed for this purpose with the aid of two optimization methods: Q-learning and Double Deep Q-Network (DDQN).
Xu \textit{et al.} \cite{xu2022maximizing} introduced a distributed distributional Deep Deterministic Policy Gradient (D4PG) approach for each RSU in the system. This approach aimed to specify sensing information, frequency, uploading priority, power of transmission, and bandwidth for the system's vehicles. The proposed model incorporated local networks and target networks, comprising a critic network and a policy network. Experiences were stored in an initialized buffer for replay, and the actions taken by the agent network aimed to maximize the quality of view through sensing while minimizing costs.

Unmanned aerial vehicles (UAVs) are AVs that are widely applied to many strategies related to sensing the environment. Wang \textit{et al.} \cite{33} devised an RL-based resource allocation strategy for a group of UAVs equipped with an ISAC system. A combination of mutual information and communication rate was formulated as the reward function. The authors proposed different algorithms, including Q-learning and deep Q-network (DQN), to learn a resource allocation policy that maximizes the reward. 

Table \ref{table:vehicles} presents a summary of the mentioned ML approaches in ISAC for AVs. Other ISAC-ML based works that considered AV systems include \cite{xu2022deep} for ISAC-assited AVs, \cite{22,graff2023deep,ahmed2022hybrid} for vehicular networks, \cite{xu2022computer,chen2021radar,lee2022intelligent} for vehicle to everything (V2X), \cite{21,wang2022cap} for vehicle to infrastructure (V2I), and \cite{charan2022towards,34} for UAVs. 

\begin{table}
\centering
\caption{Summary of key works in ML-ISAC for autonomous vehicular networks.}
\label{table:vehicles}
\resizebox{\linewidth}{!}{%
\begin{tabular}{|>{\hspace{0pt}}m{0.085\linewidth}|>{\hspace{0pt}}m{0.065\linewidth}|>{\hspace{0pt}}m{0.05\linewidth}|>{\hspace{0pt}}m{0.123\linewidth}|>{\hspace{0pt}}m{0.406\linewidth}|>{\hspace{0pt}}m{0.206\linewidth}|} 
\hline
Ref.                        & Technology                       & AI technique & Criteria                                 & Input                                                                                                                                                                                                        & Output                                                                                                                             \\ 
\hline
\cite{chu2022ai}            & Waveform design                  & RL           & 1.Duelling FCNN\par{}2.Q-learning      & Observation:~number of packets perqueue and channel quality                                                                                                                                                  & 1.Action:~Number of frames per coherent processing interval\par{}2.A mapping between the observations and the actions (Q-table)  \\ 
\hline
\cite{zhang2023deep1}       & 1.Demodulation\par{}2.Tra-cking & DL           & 1.FCNN\par{}2.YOLOv5 \cite{yolo5}      & 1.Received signal\par{}2.Range-velocity spectrum image                                                                                                                                                     & 1.Recovered symbols\par{}2.Target bounding boxes in the input image                                                              \\ 
\hline
\cite{mu2021integrated}     & Beamforming                      & DL           & FCNN                                     & Received echo                                                                                                                                                                                                & Relative angle of a target vehicle                                                                                                 \\ 
\hline
\cite{hieu2022transferable} & System design                    & RL           & 1.FCNN-based DDQN\par{}2.Q-learning    & Observation: states of the channel, the road, the weather, self speed, and the state of nearby vehicles~                                                                                                     & 1.Action: radar mode or communication mode\par{}2.Q-table                                                                        \\ 
\hline
\cite{xu2022maximizing}     & Resource allocation              & RL           & A couple of critic-action FCNNs(4 FCNNs) & State: Time slot index, RSU index, the set of distances between vehicles and the RSU, information that can be sensed by the vehicle, cached information in the RSU, and the set of views required by the RSU & Action: sensing information and frequencies, uploading priorities, and transmission power                                          \\
\hline
\end{tabular}
}
\end{table}

\subsection{Data-driven ISAC-assisted THz Communication} 

In the realm of 6G networks, THz Integrated Sensing, Communication, and Intelligence (ISCI) technology holds immense promise due to its utilization of a vibrant spectrum \cite{QADIR2022}. Sharing resources between THz ISCI counterparts can lead to increased spectrum efficiency, elimination of additional hardware costs, and improved energy efficacy \cite{wu2021thz}. Achieving these advantages relies heavily on the extent of shared resources.
THz ISCI systems are particularly inclined towards single-carrier ISAC waveforms, owing to THz channels' minimal delay spread and power amplifier efficiency. However, deploying wireless technologies operating at such high frequencies and short wavelengths poses challenges, such as limited range and increased attenuation levels. To address these challenges, researchers have explored the use of ultra-massive Multiple-Input Multiple-Output (MIMO) systems \cite{sarieddeen2019terahertz}.
In this context, the authors in  \cite{elbir2021terahertz} developed a model-based hybrid beamforming approach for ultra-massive MIMO systems operating in the low-THz band where ISAC was deployed. They proposed a cascaded combination of CNNs to predict the angles at which targets exist and to design hybrid beamformers with lower complexity than traditional model-based algorithms.

Some other works focused on designing the receiver of THz-ISAC systems using DL techniques. For instance, a multi-task neural network (NN) was formulated in \cite{wu2021thz} to develop an AI detector mechanism for sensing DFT-spread-OFDM in an ISAC system. A DL technique is implemented in \cite{wu2022sensing} to design a receiver for THz ISAC signals. Two FCNNs were designed. One FCNN was trained to predict the sensing parameters, range, and velocity, while the other FCNN was a two-level network that was trained to recover the data symbols. Both FCNNs were trained in a supervised fashion with the received data blocks in the frequency domain as inputs. A summary of the data-driven methods in THz-ISAC is  presented in Table \ref{table:thz}.



\begin{table}
\centering
\caption{Summary of key works in DL-ISAC for THz.}
\label{table:thz}
\resizebox{\linewidth}{!}{%
\begin{tabular}{|>{\hspace{0pt}}m{0.146\linewidth}|>{\hspace{0pt}}m{0.104\linewidth}|>{\hspace{0pt}}m{0.09\linewidth}|>{\hspace{0pt}}m{0.106\linewidth}|>{\hspace{0pt}}m{0.221\linewidth}|>{\hspace{0pt}}m{0.269\linewidth}|} 
\hline
Ref.                      & Application     & ML
  technique & Criterion       & Input                                   & Output                                         \\ 
\hline
\cite{elbir2021terahertz} & Beamforming     & DL             & CNN             & Received radar signal covariance matrix & Hybrid beamformer vectors                      \\ 
\hline
\cite{wu2021thz}          & Receiver design & DL             & multi-task FCNN & Received signal                         & Recovered data symbols and sensing parameters  \\ 
\hline
\cite{wu2022sensing}      & Receiver design & DL             & 2 FCNNs         & Received signal                         & Recovered data symbols and sensing parameters  \\
\hline
\end{tabular}
}
\end{table}

\subsection{Data-driven ISAC-assisted Radar Systems}

\begin{figure*}
\centering
\includegraphics[width=\textwidth,height=6cm]{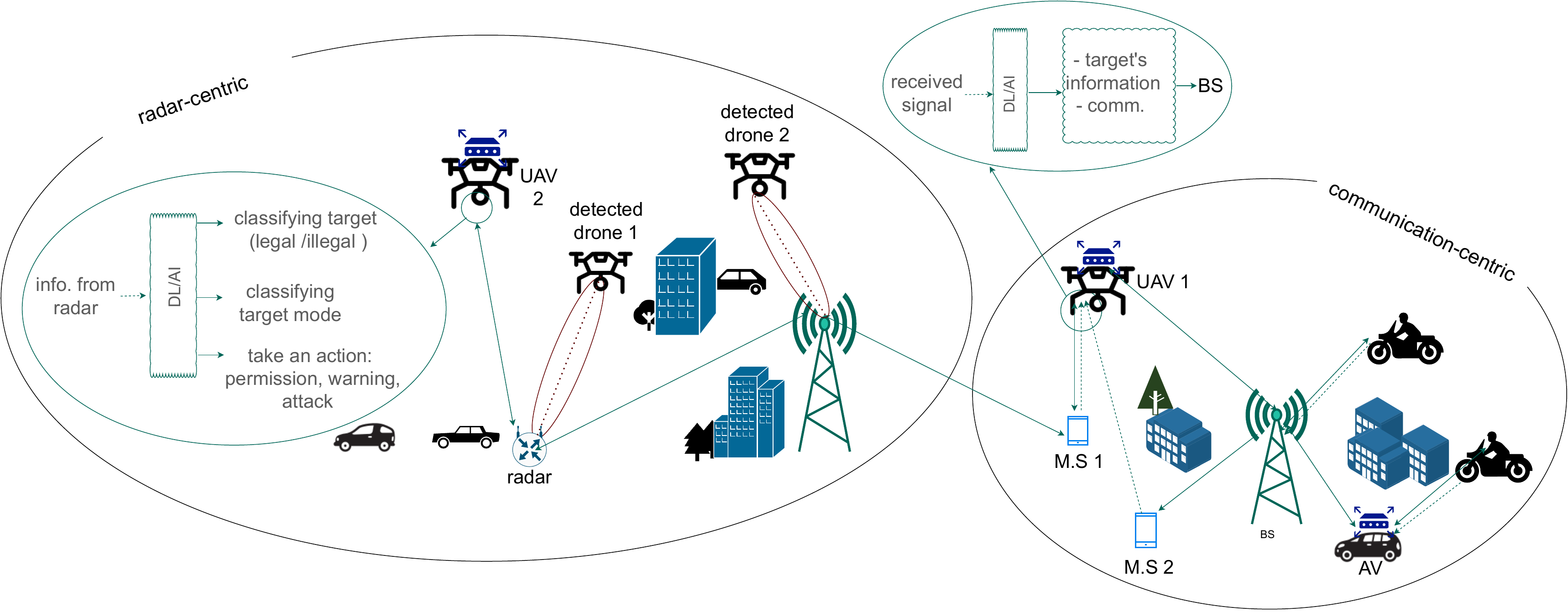}
\caption{Shows an overview diagram for both sensing-centric and communication-centric. Sensing-centric presents how communication setups can help radar function, while communication-centric shows how sensing equipment could enhance the communication performance}
\label{radar}
\end{figure*}

DL has many applications in JRC systems. Possible scenarios of both radar-centric and communication-centric JRC systems  are depicted in Fig. \ref{radar}. 
For instance, Chen \textit{et al.} \cite{chen2021radar} proposed two CNN-based DL methods to reduce the beam training overhead for link configuration in V2X systems. The authors proposed a radar-aided method that performs a radar-to-communication (R2C) translation. One model performs R2C by predicting the azimuth power spectrum (APS) of the communication link given the radar APS. The other model estimates R2C covariance columns given radar spatial covariance columns. The authors reported that the APS-based and covariance-based DNNs increased the rate by 13.3\%  and 21.9\%, respectively. V2X was also considered in \cite{lee2022intelligent}, where a resource allocation strategy based on DRL was sought. In the case of multi-agent scenarios (i.e., the existence of multiple radar-assisted vehicles), a narrow-band control channel was preserved for vehicles to communicate and coordinate their use of sensory data. This process was arranged using a graph neural network (GNN), where nodes represented the radar-assisted vehicles and the edges represented the control channel. The proposed DRL solution aimed to maximize the transmission of useful data while keeping timely radar operation through division of time and directional communication. Link configuration between road infrastructure and vehicles was discussed in \cite{graff2023deep}. The authors proposed three DNNs to convert the features from the radar domain to the communication domain. 
In \cite{xu2022deep}, an RL-based framework was used to learn a quantized transmit beamforming vector for a sparse transmit array in mobile JRC systems. 
The authors in \cite{demirhan2022radar} addressed the problem of beam alignment in radar-aided mmWave systems by a DL approach to reduce the training overhead. The radar system measurements were used to extract the range, angle, and velocity of the target via 2D and 3D Fourier transforms. The extracted information is passed through a CNN with average pooling layers. The output of the CNN indicates the beam index. 

Table \ref{table:radar} summarizes the data-driven techniques used for ISAC-assisted Radar systems. Other ML works that considered radar-assisted communication systems include \cite{zhang2023deep1,huang2021mimo20,277}.





\begin{table}
\centering
\caption{Summary of key works in ML-ISAC for radar.}
\label{table:radar}
\resizebox{\linewidth}{!}{%
\begin{tabular}{|>{\hspace{0pt}}m{0.1\linewidth}|>{\hspace{0pt}}m{0.085\linewidth}|>{\hspace{0pt}}m{0.065\linewidth}|>{\hspace{0pt}}m{0.138\linewidth}|>{\hspace{0pt}}m{0.242\linewidth}|>{\hspace{0pt}}m{0.304\linewidth}|} 
\hline
Ref.                      & Technology         & AI technique & Criterion                                        & Input                                                                         & Output                                                                                                       \\ 
\hline
\cite{chen2021radar}      & Link configuration  & DL            & 2 CNNs                                           & 1. Radar APS \par{}2. Radar spatial covariance columns                        & 1. Communication APS \par{}2. R2C covariance columns                                                         \\ 
\hline
\cite{lee2022intelligent} & Resource allocation & RL            & GNN (for multi-agent)                            & State of the local environment, local sensory data, and neighborhood links         & Action:~radar detection, data transmission, the direction of data transmission, and the null action              \\ 
\hline
\cite{graff2023deep}      & Link configuration  & DL            & 1.CNN\par{}2.6-layer FCNN\par{}3. 5-layer FCNN & 1.Radar APS\par{}2.Radar spatial eigenvector\par{}3. Radar covariance vector & 1.Communication APS\par{}2.Communication link spatial eigenvector\par{}3. Communication covariance vector  \\ 
\hline
\cite{xu2022deep}         & Beamforming         & RL            & Actor-critic networks with KNN mapping           & State of the transmit array quantized~

phase shifter                         & An action that is mapped to a certain beamformer vector                                                      \\ 
\hline
\cite{demirhan2022radar}  & Beamforming         & DL            & CNN                                              & Range, angel, and velocity of the target                                      & Beam index                                                                                                   \\
\hline
\end{tabular}
}
\end{table}

\subsection{Data-driven Approaches for Beamforming in ISAC}
The design of predictive beamforming is a crucial aspect in the realization of ISAC systems. Using DL techniques, ISAC systems can extract information from the reflected signal or the output of other sensing tools, such as a camera, to help with beamforming as shown in Fig. \ref{AVs}. Charan \textit{et al.} \cite{charan2022towards} suggested two data-driven methods (using visual data and positioning features) for quick and precise beam prediction in UAV networks. A CNN and an FCNN were the respective network architectures for the two approaches. Similarly, Xu \textit{et al.} \cite{xu2022computer} used cameras in V2X systems and proposed a vision-based DNN for optimal beamformers inference. Their DNN consisted of 1D convolutional layers for input spatial information, a transformer encoder, and fully connected layers.

The accuracy of CE or angular parameter estimation is essential for predictive beamforming. However, traditional methods suffer degraded performance due to errors in estimated historical CSI \cite{6,8,13}. In \cite{21}, the authors proposed CLRNet for ISAC-assisted vehicular networks to improve angle prediction for predictive beamforming. By combining LSTM and CNN, the CLRNet leverages spatial and temporal dependencies from historical angle estimates to facilitate precise angle prediction.
Liu \textit{et al.} \cite{22} focused on reducing the complexity of conventional predictive beamforming algorithms, imposed by CE and channel tracking, in ISAC-assisted vehicular systems. The proposed architecture is composed of CNN and LSTM modules, allowing the model to extract spatial and temporal features of the input historical CSI. An unsupervised training scheme was used to minimize the negative sum rate, where ISAC constraints were added to the loss function as penalty terms. Adhikary \textit{et al.} \cite{adhikary2023integrated} designed a holographic MIMO transceiver, where 3D beams are designed using a DL model. The DL architecture is composed of a variational autoencoder that reconstructs the estimated distances by the far-field signal, and a GRU for beamforming.

Moreover, an ISAC high-speed railway (HSR) mmWave wireless network was considered in \cite{277} by using a DRL scheme to control two mmWave beams, the communication beam, and the radar beam. The action space controls the space between the two beams and their beam widths. The proper action is predicted by a deep Q-network (DQN) given the state of the dynamic environment and previous actions. Liu \textit{et al.} \cite{liu2023distributed} proposed a DL network architecture for interference management in ISAC systems via proper power allocation. Unsupervised learning was utilized, and then transfer learning was used to predict a proper beamforming scheme for interference management.

Another DRL-based beam management scheme is proposed in \cite{288}, where user location uncertainty was considered in mmWave networks in a joint vision-aided sensing and communication system. Features were extracted from satellite images to enhance the localization accuracy. To manage beamforming for the localized UEs, a clustering method was first implemented, where a single beam is dedicated to a cluster. A DRL technique was implemented to design the corresponding beamformers depending on the clusters and channel information.


Table \ref{table:beamform} summarizes the different ML techniques utilized for beamforming design. Several other ML works considered the problem of beamforming in various ISAC settings, such as \cite{mu2021integrated, liu2022learning, hieu2022transferable, xu2022maximizing} in AV networks, \cite{xu2022deep,demirhan2022radar} in radar, and \cite{elbir2021terahertz} in radar. More details about these works are found in their respective use case subsections.






\begin{table}
\centering
\caption{Summary of key works in DL-ISAC for beamforming.}
\label{table:beamform}
\resizebox{\linewidth}{!}{%
\begin{tabular}{|>{\hspace{0pt}}m{0.094\linewidth}|>{\hspace{0pt}}m{0.115\linewidth}|>{\hspace{0pt}}m{0.058\linewidth}|>{\hspace{0pt}}m{0.19\linewidth}|>{\hspace{0pt}}m{0.277\linewidth}|>{\hspace{0pt}}m{0.2\linewidth}|} 
\hline
Ref.                          & Technology                         & AI  technique & Criteria                                                         & Input                                                                                                                                                       & Output                                                                                                                                                                  \\ 
\hline
\cite{charan2022towards}      & UAV network                        & DL            & 1. CNN\par{}2. FCNN                                              & 1. BS camera footage snapshots\par{}2.UAV's height, GPS position, and distance from the BS                                                                 & Beam index                                                                                                                                                              \\ 
\hline
\cite{xu2022computer}         & V2X                                & DL            & 1D-CNN, self-attention module, feed-forward module and an MLP    & Mobile station (MS) location and vehicle distribution features from the MS camera snapshots                                                                 & Beam index pair                                                                                                                                                         \\ 
\hline
\cite{21}                     & V2I                                & DL            & CLRNet (CNN, LSTM, FC)                                           & Historical angles                                                                                                                                           & Current angle                                                                                                                                                           \\ 
\hline
\cite{22}                     & V2I                                & DL            & CNN, LSTM units, and an FC layer                                 & Historical channel information                                                                                                                              & Predictive beamforming matrix                                                                                                                                           \\ 
\hline
\cite{adhikary2023integrated} & Holographic MIMO                   & DL            & 1.Variational autoencoder\par{}2.A gated recurrent unit        & 1. Noisy distance estimation\par{}2.Reconstructed distance vector                                                                                          & 1.Reconstructed distance vector\par{}2.3D beams and power allocation                                                                                                  \\ 
\hline
\cite{277}                    & HSR                                & RL            & CNN- based DQN                                                   & State: Train positions,~sub-6GHz SNR, mmWave SINR, historical sensing and communication beamwidths                                                          & Action: Beam separation, radar beamwidth, communication beamwidth                                                                                                       \\ 
\hline
\cite{288}                    & Generic multi-user mmWave networks & ML            & 1.FCNN\par{}2.UK-Medoids clustering\par{}3.LSTM network policy & 1. Ratio measures and pixel features extracted from satellite images\par{}2.Noisy UE locations\par{}3.State: Channel condition between the UE and the BS~ & 1. Noisy UE locations \par{}2.UE clusters, where a single beam is dedicated to every cluster\par{}3.Action: Allocating a~Resource Block Group to a user in each beam  \\
\hline
\end{tabular}
}
\end{table}

\subsection{Data-driven methods in ISAC for tracking and localization}
\label{subsec:loc}

ISAC technology has potential for improving localization and tracking in next-generation communication systems \cite{39}. Localization has been studied in various ISAC systems (e.g., \cite{16,jing2022isac,yu2022location,gao2023cooperative,dong2022localization}), but ML techniques are rarely used. A data-generation approach and a DL-supervised training scheme for localization in UAV-ISAC systems were proposed in \cite{34}. In \cite{37}, an integrated indoor-outdoor (IO) sensing and positioning scheme based on a random forest classifier was proposed. The authors used weighted $K$-nearest neighbors (KNN) for online positioning upon receiving a request from user equipment (UE). In \cite{wei2022visible}, a federated learning framework was developed to design a multi-user ISAC system, where the sensing functionality is dedicated to indoor positioning via visible light. The multi-task network shared by users was used to jointly estimate the positions and the channels. Gao \textit{et al.} \cite{gao2022toward} proposed a dataset generation method that obtains indoor positioning features in MIMO channels. The authors used the dataset to first train well-known DNNs, such as ResNet-18 \cite{he2016deep10} and GoogLeNet \cite{szegedy2015going}. They then trained a novel multi-path CNN architecture using the generated dataset, where better performance was achieved in terms of mean square error (MSE) and standard deviation. Indoor localization in ISAC systems was also considered in \cite{khunteta2022ai}. A signal processing scheme at the receiver was formulated to obtain channel information in the angular domain. Then, given the obtained information, the authors trained an FCNN to predict the total number of users in the area of operation and the number of users in every sector of the readily segmented area. 

Table \ref{table:loc} summarizes the mentioned ML approaches in the field of tracking and localization for ISAC. Other localization works discussed in other use cases include \cite{288,ahmed2022hybrid,adhikary2023integrated}.

\begin{table}
\centering
\caption{Summary of key works in DL-ISAC for tracking and localization.}
\label{table:loc}
\resizebox{\linewidth}{!}{%
\begin{tabular}{|>{\hspace{0pt}}m{0.106\linewidth}|>{\hspace{0pt}}m{0.14\linewidth}|>{\hspace{0pt}}m{0.092\linewidth}|>{\hspace{0pt}}m{0.183\linewidth}|>{\hspace{0pt}}m{0.175\linewidth}|>{\hspace{0pt}}m{0.24\linewidth}|} 
\hline
Ref.                  & Technology                           & AI~ technique      & Criteria                                  & Input                                     & Output                                                      \\ 
\hline
\cite{37}             & Cellular networks                    & ML                 & Random-forest-classifier and weighted KNN & User measurement reports                  & UE position~                                                \\ 
\hline
\cite{wei2022visible} & Indoor positioning via visible light & Federated learning & Custom RNN, LSTM units and FCNN           & Shared sparsity-aware network             & Position and the channel                                    \\ 
\hline
\cite{gao2022toward}  & Indoor positioning                   & DL                 & multi-path CNN                            & Channel frequency response                & UE position                                                 \\ 
\hline
\cite{khunteta2022ai} & Indoor positioning                   & DL                 & FCNN                                      & Channel information in the angular domain & Total number of users and number of users per area segment  \\
\hline
\end{tabular}
}
\end{table}

\subsection{Data-driven methods in ISAC for spectrum sensing}
Another evident, yet rarely acknowledged, application of ISAC technology is spectrum sensing (SS) \cite{liu2022integrated}. While some works discuss SS in the context of ML techniques (e.g., \cite{24,25,gao2019deep,zheng2020spectrum,solanki2021deep}), less work studied SS in the context of ISAC use-cases and ML technologies. For instance, Liu \textit{et al.} \cite{23} considered spectrum sensing in space-air-ground integrated networks (SAGIN) systems, where spectrum management is usually challenging. The authors proposed a data-driven approach for SS to eliminate the uncertainty associated with noise and pinpoint the channels occupied by the primary transmission. A CNN classifier was utilized to predict channel occupancy. A similar binary classification objective was approached by \cite{soni2022pu}, but with a different DL architecture (c.f., Table \ref{table:spec}). Then, Ahmed \textit{et al.} \cite{ahmed2022hybrid} proposed a joint framework for spectrum sensing and localization in Internet of Vehicle (IoV) networks. They proposed a CNN-based model with skip connection layers and an Atrous Spatial Pyramid Pooling (ASPP) \cite{chen2017deeplab} module. The CNN is trained to classify the spectrogram of the received signal whether it is a noise or a primary signal. Based on the network's decision, a support vector machine (SVM) was used for localization. 
Table \ref{table:spec} summarizes various data-driven approaches in ISAC for spectrum sensing applications.

\begin{table}
\centering
\caption{Summary of key works in DL-ISAC for spectrum sensing.}
\label{table:spec}
\resizebox{\linewidth}{!}{%
\begin{tabular}{|>{\hspace{0pt}}m{0.117\linewidth}|>{\hspace{0pt}}m{0.083\linewidth}|>{\hspace{0pt}}m{0.075\linewidth}|>{\hspace{0pt}}m{0.308\linewidth}|>{\hspace{0pt}}m{0.154\linewidth}|>{\hspace{0pt}}m{0.202\linewidth}|} 
\hline
Ref.                   & Technology      & AI technique & Criteria                                                                   & Input                             & Output                                                                                                                      \\ 
\hline
\cite{23}              & SAGIN           & DL            & CNN                                                                        & Received signal covariance matrix & Class: channel occupancy assumptions\par{}$H_0$:No primary signal\par{}$H_1$:Primary signal                               \\ 
\hline
\cite{soni2022pu}      & Cognitive radio & DL            & A multi-layer perceptron combined with an input propagation trainable path & Signal power                      & Class: channel occupancy assumptions\par{}$H_0$:The null hypothesis (Noise)\par{}$H_1$:Primary signal                         \\ 
\hline
\cite{ahmed2022hybrid} & IoV             & DL            & CNN, skip connections, and ASPP module                                     & Received signal spectrogram       & Class: channel occupancy assumptions\par{}- Spectrum hole (no transmission)\par{}- No spectrum hole (primary transmission)  \\
\hline
\end{tabular}
}
\end{table}

\subsection{Data-driven approaches in ISAC-assisted edge computing}
Edge computing (EC) plays important roles in modern and future ISAC-enabled networks, such as reducing heavy computational requirements from the user devices when DL/DRL algorithms are employed \cite{ding2022joint,wang2022noma}. Few works focused on the use of DL and DRL for systems that employ ISAC and edge computing. Yuan \textit{et al.} \cite{yuan2020cross} considered smart road systems that enable vehicle operations via capabilities that can be viewed in two integrated domains, transportation domain (e.g., road structure) and information domain (e.g., communication, sensing, and computing). The authors proposed a DRL approach to optimally allocate resources from the cross-domain (e.g., data transmission, path planning, computation offloading). The multi-agent framework is based on CNNs that analyze the traffic state to predict the optimum action. Souza \textit{et al.} \cite{de2020enhancing} proposed a path planning framework for self-driving vehicles with the capabilities of performing mobile EC (MEC) and ISAC. In the proposed framework, MEC aims to predict congestion points using an LSTM-based architecture. Given such predictions, road structure (represented by a graph), and the starting and end points, a Q-learning algorithm was proposed to formulate a Q-table of the most reliable path.

The complexity and the dynamic of road environments render autonomous driving offline-trained DL systems ungeneralizable and prone to the curse of dimensionality. Such issues were addressed in \cite{wu2022intelligence}. The authors proposed a multi-access EC (MAEC) framework to train a group of DNNs, each of which is suited to a specific road segment. The models are trained by MAEC nodes via a blockchain system. Table \ref{table:edge} summarizes various data-driven techniques and models used for ISAC-assisted EC works.

\begin{table}
\centering
\caption{Summary of key works in DL-ISAC for edge computing.}
\label{table:edge}
\resizebox{\linewidth}{!}{%
\begin{tabular}{|>{\hspace{0pt}}m{0.102\linewidth}|>{\hspace{0pt}}m{0.077\linewidth}|>{\hspace{0pt}}m{0.069\linewidth}|>{\hspace{0pt}}m{0.181\linewidth}|>{\hspace{0pt}}m{0.344\linewidth}|>{\hspace{0pt}}m{0.165\linewidth}|} 
\hline
Ref.                   & Application   &  AI Technique    & Criteria                                                                      & Input                                                                                                                                                      & Output                                                                         \\ 
\hline
\cite{yuan2020cross}   & Smart roads  & RL               & 1.CNN-based value iteration network (VIN)\par{}2. CNN-based two-branched DQN & 1. Observation: destination and current traffic status\par{}2.Observation: origin-destination, traffic status, host RSU server, and edge computing loads & 1.Action: Route selection\par{}2.Action: Agent migration and route planning  \\ 
\hline
\cite{de2020enhancing} & ISAC-MEC-AVs & 1.DL\par{}2.RL & 1.LSTM-based RNN (representing MEC)\par{}2.Q-learning                       & Observation: congestion state (RNN prediction) at a certain time stamp, road graph, urban dynamics                                                         & Q-table: observation-routing pairs                                             \\
\hline
\end{tabular}
}
\end{table}
\begin{figure}[!t]
\centering
\includegraphics[width=90mm]{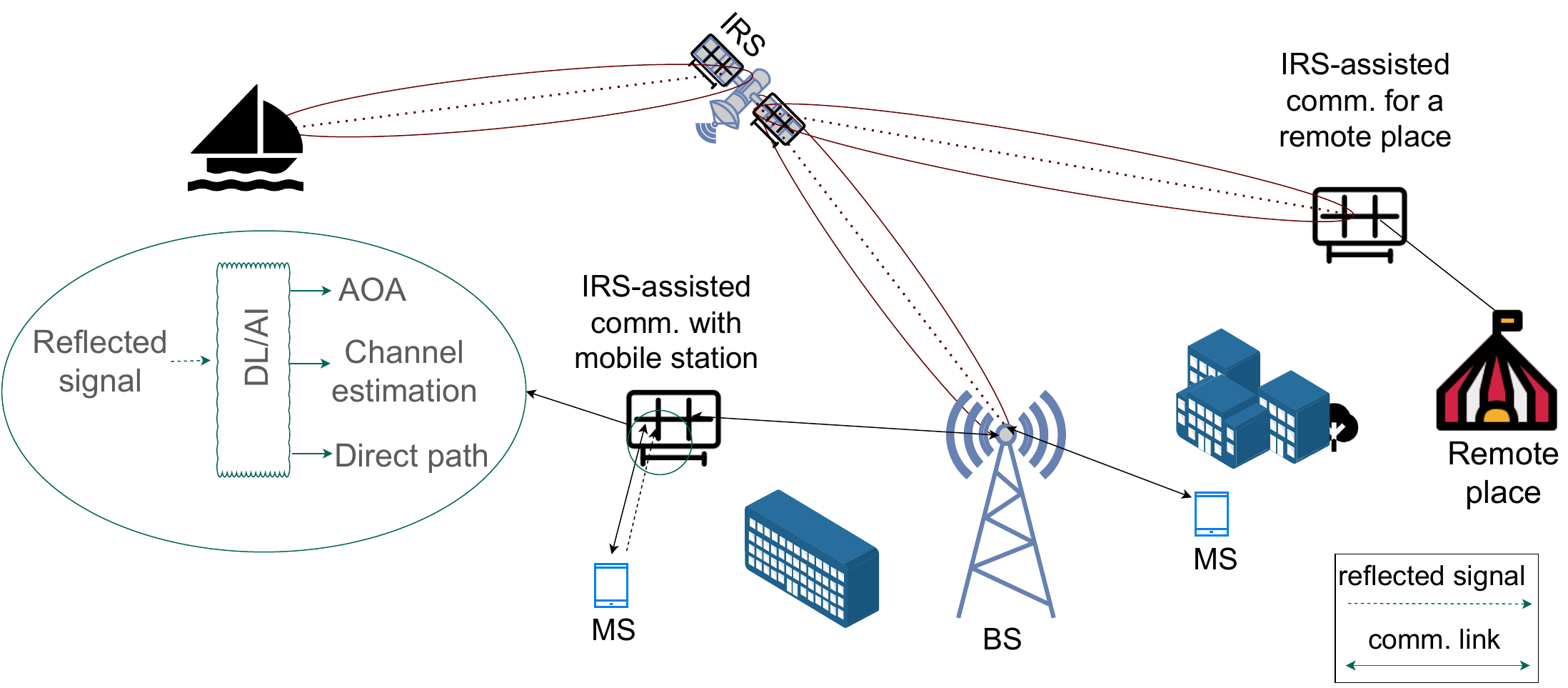}
\caption{Schematic representation of an IRS-assisted ISAC system.} 
\label{IRS}
\end{figure}

\subsection{Data-driven methods for channel estimation and IRS}

Efficient CE is crucial for characterizing received symbols in wireless systems with time-varying and/or frequency-selective channels. Huang \textit{et al.} \cite{huang2021mimo20} proposed a MIMO radar-based CE framework that divides into two stages: AoA/AoDs estimation at the radar module and gain estimation at the communication module. arrival/departure (AoA/AoDs) were obtained using a model-based algorithm, while gains were obtained using a residual denoising autoencoder (RDAE). The RDAE was trained to denoise input signals, where the least squares (LS) method is applied to the output for gain estimation.

On the other hand, IRS is a physical surface with reflecting elements that are set up in propagation environments to overcome the coverage holes that occur in wireless communication \cite{Javaid2023}. Fig. \ref{IRS} shows examples for IRS deployment in different propagation settings. Dynamic V2I scenarios in IRS-assisted ISAC systems were considered by Wang \textit{et al.} \cite{wang2022cap}. To estimate AoAs at both the base station and deployed IRS, the authors proposed CAP-Net, a DNN consisting of convolutional layers and LSTM units. CAP-Net was trained using historical covariance data of received echo to predict AoAs.

CE becomes critical in IRS-assisted systems as passive IRS cannot perform signal processing. Therefore, Zhang \textit{et al.} \cite{zhang2023self} presented a self-supervised learning approach to IRS-CE. During training, a DNN learns to output signals, similar to the original, when provided with the noisy version of the signal. The trained denoiser is used to improve channel estimation. The approach is shown to be suitable for various communication systems, including ISAC. 

CE in multiple-input-single-output (MISO) IRS-assisted ISAC systems was considered in \cite{liu2022deep}. The problem was divided into three stages: estimating direct SAC channels in the first stage, individually reflecting communication and sensing channels in the second and third stages, respectively. To accommodate the inherent propagation differences between direct and reflected channels in a full-duplex ISAC base station, two CNNs were designed and trained to make predictions at all three stages. Table \ref{table:irs} summarizes the works using DL-ISAC techniques in for IRS/CE.



\begin{table}
\centering
\caption{Summary of key works in DL-ISAC for channel estimation and IRS.}
\label{table:irs}
\resizebox{\linewidth}{!}{%
\begin{tabular}{|>{\hspace{0pt}}m{0.1\linewidth}|>{\hspace{0pt}}m{0.112\linewidth}|>{\hspace{0pt}}m{0.069\linewidth}|>{\hspace{0pt}}m{0.165\linewidth}|>{\hspace{0pt}}m{0.19\linewidth}|>{\hspace{0pt}}m{0.298\linewidth}|} 
\hline
Ref.                 & Technology             & AI technique & Criteria                                                                    & Input                                                                           & Output                                                                     \\ 
\hline
\cite{wang2022cap}   & IRS-assisted V2I       & DL            & CAP-Net (conv layers and LSTM)                                              & Historical covariance of the received echo                                      & AoAs                                                                       \\ 
\hline
\cite{zhang2023self} & General applications   & DL            & CNN (Trained as a denoiser)                                                 & Channel estimated by LS method                                                  & Channel estimation                                                         \\ 
\hline
\cite{liu2022deep}   & IRS-assisted ISAC MISO & DL            & 1.Direct estimation CNN (DE-CNN)\par{}2.Reflected estimation CNN (RE-CNN) & 1.Received direct signals\par{}2.Total received signals and DE-CNN estimation & Direct (DE-CNN) and reflected (RE-CNN) sensing and communication channels  \\
\hline
\end{tabular}
}
\end{table}

\subsection{Data-driven methods for other ISAC applications}
Some DL-ISAC-oriented works cannot be grouped into one of the previous categories because they are rather focused on dispersed yet important 
applications.

\noindent\textbf{Gesture recognition:} Qi \textit{et al.} \cite{qi2023resource} proposed a federated transfer learning framework for gesture recognition with Wi-Fi sensing as an indoor deployment for ISAC.

\noindent\textbf{Signal design:} Xie \textit{et al.} \cite{xie2022intelligent} trained an autoencoder, in an unsupervised fashion, to extract the features of an input ISAC signal. Thus, the trained encoder is expected to fuse the ISAC signal into a lower-dimensional signal before broadcasting.

\noindent\textbf{Hardware impairments:} The problem of hardware impairments' effect on beamforming in ISAC systems when standard mathematical modeling is used (i.e., standard model-based methods) was addressed in \cite{mateos2022model}. The authors proposed an MDL approach, where some parameters of the mathematical framework are learned. The learning process adapts with hardware impairments, such as inaccurate antenna spacing, resulting in better results than the standard model-based approaches', where mathematical models are derived by assuming perfect hardware design. The authors also designed an FCNN to directly obtain the target parameters and compared the three approaches (i.e., the standard model-based, the MDL, and the data-driven approaches).

\noindent\textbf{Atmospheric sensing:} In \cite{wedage2021climate}, climate change sensing using THz communication was discussed. The advantage of environmental gases absorption by the THz signal was utilized. ML techniques were employed to analyze the power spectral density and signal path loss to determine the quantity of various gases that have an impact on the climate.

\noindent\textbf{Agriculture:} A result-oriented literature review presented in \cite{usman2022terahertz} focused on the study of vital resource preservation in precision farming field to enhance the production and harvesting of agriculture through 6G smart management systems. Nano sensors are considered for such infrastructure so that early-stage plant health can be monitored. Through a designated line of communication, the gathered information is uploaded to a distant cloud server. One use of THz communication is to establish active communication links between nano-sensors. The pioneering work on in nano-network communication was presented by \cite{kaushik2021optimization}.

\section{key challenges}
\label{section:challenges}
Data-driven approaches have shown promise compared to model-driven methods, but their implementation in ISAC systems comes with key challenges. In the following, we discuss some of these challenges.

\subsection{Limited data availability}
\label{subsec:availability}
One of the main challenges in data-driven approaches is data availability. This is particularly true in ISAC systems due to the sensitivity or confidentiality of the data, making it difficult to be collected and used \cite{buzzi2019using}. Additionally, the data collection process can be considerably expensive and time-consuming. This constraint can limit data availability and potentially lead to incomplete or biased outcomes.\cite{22}. This makes it challenging to build accurate models that can generalize effectively and prevent selective bias.

\subsection{Data quality and reliability}
Data quality and reliability can vary due to factors like sensor noise, measurement errors, and data transmission loss. This can create inaccurate or biased models that do not represent the underlying system behavior \cite{dahrouj2021overview}. Therefore, selected data must be pre-processed to remove outliers and missing values, ensuring it's in a format understood by the DL algorithm. 

\subsection{Model complexity}
The development of accurate models for ISAC systems can be challenging due to their inherent complexity, which encompasses multiple sensors, communication channels, and system components. While DL models have shown great promise in accurately predicting the behavior of systems in different scenarios, they come with computational complexity, particularly during the training phase \cite{barneto2021full}. Consequently, building models that capture the system complexity requires significant resources and time investment. Additionally, the complexity of the models can lead to overfitting, where the model performs well on the training data but poorly on new, unseen data. Therefore, developing models that balance the complexity of the DL models with their predictive capabilities remains an ongoing challenge.

\subsection{Limited interpretability}
Interpreting data-driven approaches can be challenging, particularly when dealing with complex models, which can limit the understanding of the underlying mechanisms driving the system's behavior. This can make it difficult to identify and address any issues that may arise \cite{33,37}, highlighting the importance of developing interpretable models for informed decision-making and improved system performance.


\subsection{Integration with traditional methods}
Although data-driven approaches have shown promise, integrating them with traditional methods such as physics-based models or heuristics may be necessary to achieve optimal results. However, integrating these methods can be challenging due to differences in data types, assumptions, and algorithms. To develop accurate and reliable models, it is crucial to balance data-driven and traditional methods appropriately, depending on the specific application and available resources, while combining the strengths of both approaches \cite{liu2021deep}.  Nonetheless, some classical methods may not be suitable for integrating with data-driven models \cite{zheng2019intelligent}.  

\subsection{Privacy and security concerns}
As discussed in \ref{subsec:availability}, ISAC systems have the potential to collect sensitive information, including personal health data and location information. Therefore, it is imperative to collect and use this data in a way that safeguards the privacy and security of individuals \cite{25}.


\section{Future research directions}
\label{section:future}

ML algorithms show promise for accelerating the adoption of ISAC technologies, but there are still gaps and unanswered questions that require further research. This section highlights some of these difficulties and gaps to aid researchers in filling them.

\subsection{Algorithms based on less amount of data}
Although recent advancements in ML algorithms have shown significant progress, it is still necessary to utilize vast amounts of data to achieve meaningful results. Fortunately, sensor fusion techniques \cite{wild2021joint} and IRS provide potential solutions for this problem, particularly in non-line-of-sight scenarios, by utilizing various data sources. For example, outdoor data can be used for localization and tracking, enhancing personalized services. Moreover, real-world data on traffic flows can be combined with sensory data gathered by AVs, further improving performance. Maps, on the other hand, represent a reliable data source that can significantly enhance ML algorithms in the context of AVs and intelligent roads \cite{lee20226g}.

\subsection{Handling time-varying systems}
Highly time-varying models require advanced ML approaches. Channel estimation errors increase as channels change with time more quickly \cite{liu2022survey}. Improved ML algorithms are needed to estimate channel states in such scenarios and improve ISAC performance across different mobility modes.

\subsection{Privacy and security}
When developing appropriate data sharing and access control mechanisms, it's crucial for researchers to carefully consider data privacy and security policies. The development of ISAC introduces new security challenges, such as susceptibility to manipulation, eavesdropping, and jamming attacks \cite{furqan2021wireless}. These security weaknesses may cause apprehension from using ISAC in many applications. Therefore, it's imperative to devote a significant amount of effort toward proposing effective ML/AI methods that can handle these security challenges and ensure the secure operation of ISAC.
 
\subsection{Training time issue}
Training is a crucial yet a time-consuming issue when using large data sets. Novel techniques with low latency and proper training are needed to achieve required performance, especially for time-sensitive systems like communication \cite{khalil2021deep}. Distributed deep learning algorithms and algorithms with acceptable performance using less data are two potential solutions \cite{dai2019bigdl}.

\subsection{Integrating data-based and model-based algorithms}
Creating a new architecture for each component of ISAC systems is not always necessary for ISAC development. Certain elements of the communication and sensing systems can still be utilized with some modifications. To expedite the integration of ML algorithms in ISAC and facilitate their adoption, it is crucial to develop novel approaches that enable the effective integration of data-based algorithms with model-based algorithms. Such methods would need to be tailored to the specific requirements of ISAC technology to offer improved efficiency and accuracy in the systems.

\subsection{Complexity issue}
ML algorithms have an advantage over model-based algorithms for their ability to handle complex systems without explicit modeling. As ISAC system complexity increases, more advanced AI algorithms are needed. DL and NN algorithms are promising solutions for this challenge. However, further studies are necessary to demonstrate the practical characteristics of ML algorithms in ISAC. This will enable the development of advanced algorithms that can be effectively deployed in various applications to meet the demands of complex systems \cite{kim2022survey}.


\subsection{Exploring novel ISAC use cases}
The use of ML in various ISAC applications has yet to be explored, such as automatic modulation recognition (AMR) for received signals. AMR can enhance ISAC's sensing module, e.g., spectrum monitoring \cite{zhou2023semi11}. However, there is little work using DL for AMR in ISAC systems \cite{zhang2022amr}. This is an interesting but relatively unexplored area for DL deployment in ISAC. Another example is the ISAC uplink design. In literature, few works have focused on designing and/or analyzing ISAC uplink systems \cite{wang2022noma,he2022full,liu2022evolution,ouyang2022performance,ouyang2022performance2}, and no work has been done to design the uplink using ML techniques.

The use cases in Section \ref{section:usecases} could be further explored. EC works, for example, only proposed generic ML techniques in certain AV systems. Also, novel configurations can be proposed for AV systems, where a certain application (e.g., beamforming, CE, EC) can be targeted. An example of such configurations is shown in Fig. \ref{radar}.

\section{Conclusion}
\label{section:conc}

This work highlights the role of data-driven ISAC systems in 6G and beyond. It shows that ML algorithms can improve ISAC performance and reduce costs and complexity. It reviews various deep learning applications for ISAC in domains such as autonomous vehicles, THz communication, radar systems, beamforming, tracking and localization, spectrum sensing, channel estimation, and more. It also discusses the challenges and future directions of applying ML algorithms in ISAC. This work demonstrates the importance and potential of ISAC in next-generation wireless communication systems and provides a solid basis for further research and innovation.

\bibliographystyle{IEEEtran}
\bibliography{IEEEabrv,main_ref_abr}

\vfill

\end{document}